\documentclass{IEEElmag}

\usepackage[colorlinks,urlcolor=blue,linkcolor=blue,citecolor=blue]{hyperref}

\usepackage[hyphenbreaks]{breakurl}

\usepackage{color,array}

\usepackage{graphicx}

\usepackage{CJKutf8}

\jvol{XX}
\jnum{XX}
\paper{1234567}
\pubyear{2022}
\receiveddate{xxxx 00, 0000}
\reviseddate{xxxx 00, 0000}
\publisheddate{xxxx 00, 0000}
\currentdate{xxxx 00, 0000  (Dates will be inserted by IEEE; ``published'' is the date the accepted preprint is posted on IEEE Xplore\textsuperscript{\textregistered}; ``current version'' is the date the typeset version is posted on Xplore\textsuperscript{\textregistered})}
\doiinfo{LMAG.2020.Doi Number}

\usepackage{textcomp}
\usepackage{amsmath,amsfonts,amssymb}
\usepackage{gensymb}
\usepackage[capitalise]{cleveref}

\usepackage{tikz}
\newcommand*\circled[1]{\tikz[baseline=(char.base)]{\node[shape=circle,draw,inner sep=1pt,fill=gray!10] (char) {#1};}}

\usepackage{physics}

\usepackage{orcidlink}
\begin{document}


\sptitle{Nanomagnetics}
\title{The impact of sample insulation on estimating the heating power of magnetic nanoparticles by AC calorimetry}
\author{$\text{L}$ise G. Hanson \affilmark{1} \orcidlink{0000-0002-8710-2610}}
\author{Bianca L. Hansen \affilmark{1} \orcidlink{0000-0002-4279-780X}}
\author{Thomas Veile \affilmark{1}
\orcidlink{0000-0001-5335-3502}}
\author{Mathias Zambach \affilmark{1}\orcidlink{0000-0001-9771-4613}}
\author{Niels B. Christensen \affilmark{1}\orcidlink{0000-0001-6443-2142}}
\author{Cathrine Frandsen \affilmark{1}\orcidlink{0000-0001-5006-924X}}
\affil{Department of Physics, Technical University of Denmark, 2800 Kongens Lyngby, Denmark}
\corresp{Corresponding author: Cathrine Frandsen (cfra@dtu.dk).}
\markboth{Preparation of Papers for \emph{IEEE Magnetics Letters}}{Lise G. Hanson}

\begin{abstract} 
Correct estimation of the heating power of magnetic nanoparticles is important for magnetic hyperthermia treatment. This work investigates the impact of sample insulation in AC calorimetry. We show that temperature increase in the insulation can lead to systematic errors when estimating the heating power by the corrected slope method. The errors arise if the temperature of the sample environment is kept fixed at its initial temperature in the data analysis. To correct for this, we propose the use of a local temperature difference between the sample and the sample environment.
\end{abstract}


\maketitle

\section{INTRODUCTION}\label{introduction}
Magnetic nanoparticles can act as heat sources when an alternating magnetic field is applied. This is being explored for a type of localized cancer treatment, magnetic hyperthermia, where magnetic nanoparticles are concentrated in the cancerous tissue and subsequently heated by a magnetic field {[}Gilchrist 1957, Maier-Hauff 2011, Chang 2018{]}.
Despite promising clinical results, it remains a challenge to establish measurement protocols that can reliably quantify the particle heating {[}Wildeboer 2014, Wells 2021{]}.

\begin{figure}
    \centering
\includegraphics[width=0.6\linewidth]{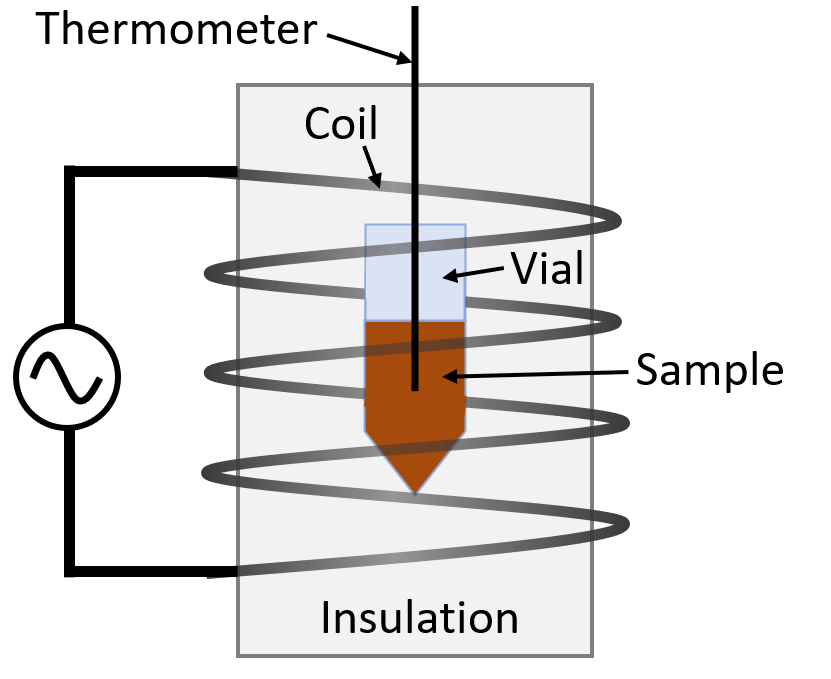}
    \caption{Illustration of a typical non-adiabatic AC calorimetry setup.}    \label{fig:NonAdiabaticCalorimetricSetup}
\end{figure}

The most common method for estimating the heating power of magnetic nanoparticles is non-adiabatic AC calorimetry {[}Andreu 2013, Fratila 2019{]}.
\cref{fig:NonAdiabaticCalorimetricSetup} illustrates a typical setup. 
A sample of magnetic nanoparticles dispersed in a liquid or cell medium is held in a vial. An alternating current (AC) in a surrounding coil generates the  magnetic field. The coil is usually water-cooled to avoid external heating of the sample. The temperature of the sample versus time is recorded by a thermometer, which must be insensitive to the AC field. The term \textit{non-adiabatic} refers to the heat loss from the sample being non-negligible. Sample insulation is often used to minimize heat loss, but there is no established convention on whether to insulate the sample or not. Solutions span from vacuum vessels to polystyrene foam shields and ambient air exposure {[}Andreu 2013{]}.

Wells {[}2021{]} showed that measurements of the heating power of magnetic nanoparticles by non-adiabatic AC calorimetry suffer from large systematic errors. This was found from comparing measurements by 17 European state-of-the-art laboratories on samples from the same batch. All laboratories except one used non-adiabatic AC calorimetry. The systematic errors inhibit the comparison of heating powers between laboratories and complicates the optimization of nanoparticles for magnetic hyperthermia. 
Alternative methods are adiabatic AC calorimetry  and AC magnetometry, but these setups are complicated to build and not widely available 
{[}Fratila 2019, Garaio 2014, Natividad 2008{]}.
Due to the prevalence and relative simplicity of non-adiabatic AC calorimetry, it is desirable to investigate the origin of systematic errors. 

In several studies, errors arising from field inhomogeneity, thermometer position, sample volume, vial material, sample geometry, and time delays have been investigated [Huang 2012, Wells 2021, Wildeboer 2014, Wang 2013]. The influence of sample insulation, however, has not been investigated systematically. 
In this letter, we show how non-linear behavior, which has not previously been accounted for, occurs in measurements, where only the sample environment has been varied. Moreover, from temperature measurements inside and outside the sample, a significant temperature increase in the sample insulation is observed. Based on this, we propose a modification of the corrected slope method taking into account the local temperature increase of the sample environment. 


\section{THEORY}
In AC calorimetry, the heating power of magnetic nanoparticles, $P_\text{MNP}$, is estimated by the temperature change in the sample. In the non-adiabatic case, the governing equations include heat transport, which happens through conduction, convection, and radiation. 
It is usually described by the lumped heat capacity model {[}Andreu 2013{]}
\begin{gather}
    C \frac{d T_s}{dt} = P_\text{MNP}-P_\text{loss},\label{eq:LumpedHeatCapacityModel}
\end{gather}
where $C$ is the heat capacity of the sample, $T_s$ is the time-dependent sample temperature, $t$ is time, $P_\text{MNP}$ is the heating power generated in the magnetic nanoparticles, and $P_\text{loss}$ is the power loss from the sample to the environment. The model is valid under the assumption that the sample temperature is sufficiently homogeneous.
Equation \eqref{eq:LumpedHeatCapacityModel} is the underlying workhorse of all non-adiabatic AC calorimetry for magnetic hyperthermia. 
Two common methods for estimating the heating power called \textit{the initial slope method} and \textit{the corrected slope method} are derived from \eqref{eq:LumpedHeatCapacityModel}. 

\subsection{Initial slope method}
\begin{figure}
    \centering
    \includegraphics[width=0.95\linewidth]{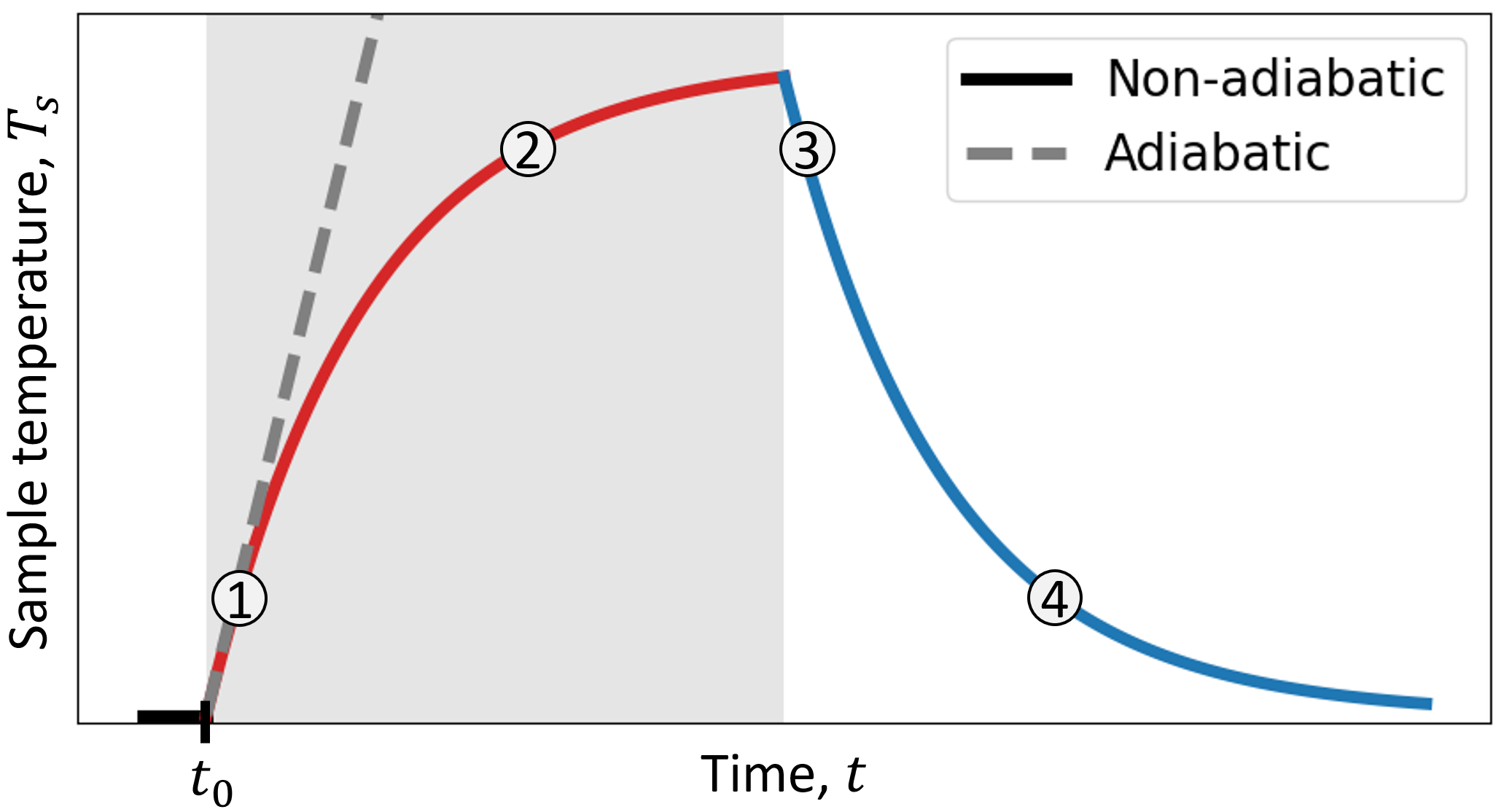}
    \caption{Sample temperature before, under, and after field application in a non-adiabatic (solid) and an adiabatic (dashed) AC calorimetric setup. The gray area indicates when the AC field is applied. The encircled numbers 1-4 denote different measurement regions.}
    \label{fig:TsVst}
\end{figure}
At the onset of the heating, the temperature difference between the sample and the surroundings is small. For a short time interval, the power loss, $P_\text{loss}$, is expected to be negligible (i.e., essentially adiabatic conditions). In this case, $P_\text{MNP}$ can be determined by
\begin{gather}
    P_\text{MNP} = C\frac{d T_s}{dt}\Big|_{t\approx 0}.
\end{gather}
Typical heating curves for non-adiabatic and adiabatic setups are illustrated in \cref{fig:TsVst}. Initially, the two curves overlap and have the same slope. The initial slope method is known to depend on the time interval in which the initial slope is determined and to systematically underestimate $P_\text{MNP}$ {[}Natividad 2009, Wildeboer 2014{]}.

\subsection{Corrected slope method}
The corrected slope method is considered the best-practice method by 
Wildeboer {[}2014{]} and Wells {[}2021{]}.
The method assumes a linear power loss, i.e., Newton's law of cooling
\begin{gather}
    P_\text{loss} = L \Delta T, \quad \Delta T = T_s - T_0,
    \label{eq:L}
\end{gather}
where $L$ is the linear loss parameter and $\Delta T$ is the temperature difference between the sample, $T_\text{s}$, and its surroundings, $T_0$. 
By combining \eqref{eq:LumpedHeatCapacityModel} and \eqref{eq:L}, the heating power can then be estimated by the corrected slope method
\begin{gather}
     P_\text{MNP}=C\frac{dT_s}{dt} + L\Delta T.
     \label{eq:CorrectedSlope}
\end{gather}
$L$ is obtained from a linear fit to a slope curve (plot of $C\, dT_s/dt$ vs. $\Delta T$) and should not depend on whether it is found from the heating or the cooling phase. The corrected slope method is described more in detail by Wildeboer [2014], who originally proposed the method. Other methods are: Box-Lucas, decay method, and steady state method. These all build on \eqref{eq:CorrectedSlope} and thus require a linear power loss {[}Wildeboer 2014, Andreu 2013{]}. 

Newton's law of cooling originates from Newton's work on forced convection {[}Winterton 1999{]}.
To which extent $P_\text{loss}$ is linearly dependent on $\Delta T$ relies on the cause of heat transfer and thus also on the sample environment. Power losses due to conduction and forced convection are expected to depend linearly on $\Delta T$, while power losses due to natural convection and radiation have a non-linear $\Delta T$-dependence  {[}Carslaw 1959{]}.

In AC calorimetry studies, where the corrected slope method is used, it is common procedure to set the temperature of the surroundings, $T_0$, equal to the initial sample temperature. $T_s(t_0)$, i.e. equal to the thermal equilibrium temperature of the sample at time $t_0$ when the field is switched on {[}Wells 2021, Wildeboer 2014{]}.
We emphasize here, in agreement with Newton's law of cooling, that heat losses occur due to the temperature difference between the sample and its surroundings. If the sample surroundings absorb heat, then the temperature of the surroundings can increase with time. Hence, the use of $\Delta T=T_s - T_0$ with $T_0$ set equal to the initial system temperature, $T_s(t_0)$, is questionable.

\subsection{Slope curves}
\begin{figure}[t]
    \centering
    \includegraphics[width=0.95\linewidth]{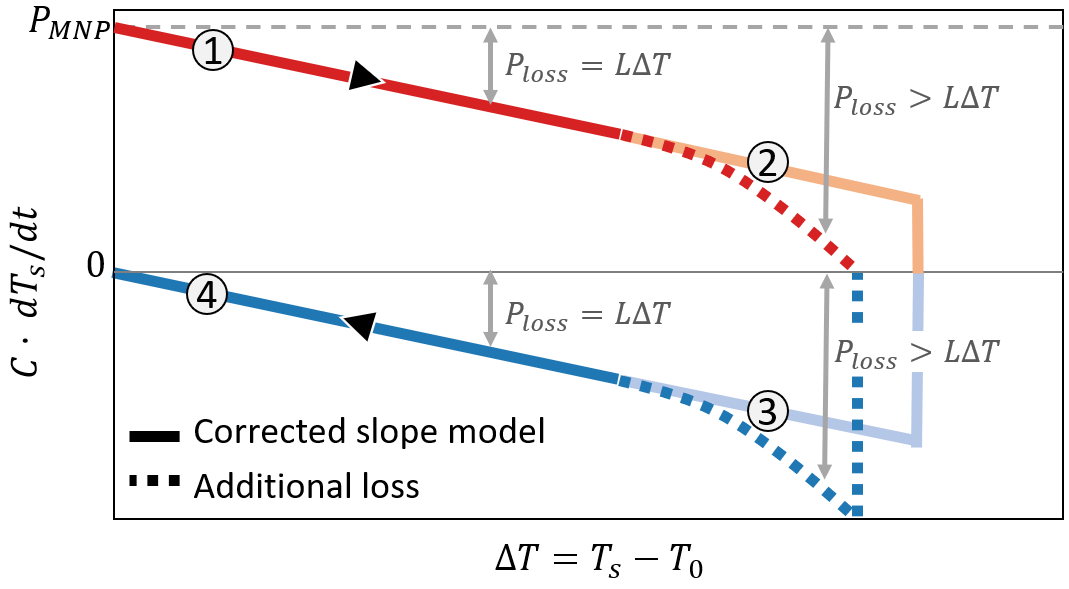}
    \caption{Theoretical slope curve. 
    Red and blue indicate the heating and cooling phase, respectively.
    The encircled numbers 1-4 are used for comparison with \cref{fig:TsVst}. The black arrows indicate the direction of time. The gray arrows indicate estimations of $P_\text{loss}$ when $P_\text{MNP}$ is constant.}
    \label{fig:SlopeCurveTheory}
\end{figure}
In this section, we introduce the concept of slope curves, which are plots of  $C\,dT_s/dt$ vs. $\Delta T$. These curves are an important tool for understanding the mechanisms of power losses, cf. \eqref{eq:LumpedHeatCapacityModel} and \eqref{eq:CorrectedSlope}.
In \cref{fig:SlopeCurveTheory}, a theoretical slope curve is illustrated. The red and blue curves correspond to the heating and cooling phase, respectively.
Equation \eqref{eq:CorrectedSlope} is directly plotted as the solid curves. The dashed lines indicate the course of the slope curve if power losses additional to $L\Delta T$, e.g. due to radiation, appear at high temperature differences. 

A comparison between the regions \circled{1}-\circled{4} in Figs. \ref{fig:TsVst} and \ref{fig:SlopeCurveTheory} gives an intuitive understanding of the slope curves. At \circled{1}, shortly after the applied field is turned on, $\Delta T$ is small and consequently the power loss is also small. This causes a rapid temperature increase i.e., a high $dT_s/dt$ value. Conversely, at \circled{2}, $\Delta T$ is high, and $dT_s/dt$ becomes small due to a high power loss. 
At \circled{3}, $\Delta T$ and the power loss remain high, but with the field turned off, $dT_s/dt$ becomes  high and negative. Finally, at \circled{4}, the initial equilibrium state is slowly approached causing $\Delta T$ and $dT_s/dt$ to be small. 

Considering \eqref{eq:LumpedHeatCapacityModel}, the power loss can be read off from the slope curves in \cref{fig:SlopeCurveTheory} at any given value of $\Delta T$. During heating 
\begin{gather}
    P_\text{loss} = P_\text{MNP}-C\frac{dT_s}{dt},
    \label{eq:PlossP}
\end{gather}
and during cooling, where $P_\text{MNP}=0$,  
\begin{gather}
    P_\text{loss}=-C\frac{dT_s}{dt}.
    \label{eq:Ploss0}
\end{gather}
The gray arrows in \cref{fig:SlopeCurveTheory} illustrate
the power loss at two different values of $\Delta T$. 
In cases with additional non-linear losses (i.e. when $P_\text{loss} >L\Delta T$), the slope curves bend downwards with increasing $\Delta T$  in both the heating and cooling phases, as shown in \cref{fig:SlopeCurveTheory}, but the power loss in the heating and cooling phase is expected to be identical at the same value of $\Delta T$.  
$P_\text{MNP}$ can depend on $T_s$ {[}Papadopoulos 2020{]}
and thus the direct reading of $P_\text{loss}$ from the heating phase of the slope curves should be done with care.

$P_\text{MNP}$ can be read from the intercept with the $y$-axis of the linear slope curve in the heating phase. Fundamentally, the corrected slope method corresponds to a linear fit to the heating phase of the slope curve and subsequent identification of $P_\text{MNP}$ as the value at $\Delta T = 0$ by an extrapolation of this fit. 

\section{STATE-OF-THE-ART AND NON-LINEARITY}
\begin{figure}[b]
    \centering
    \includegraphics[width=\linewidth]{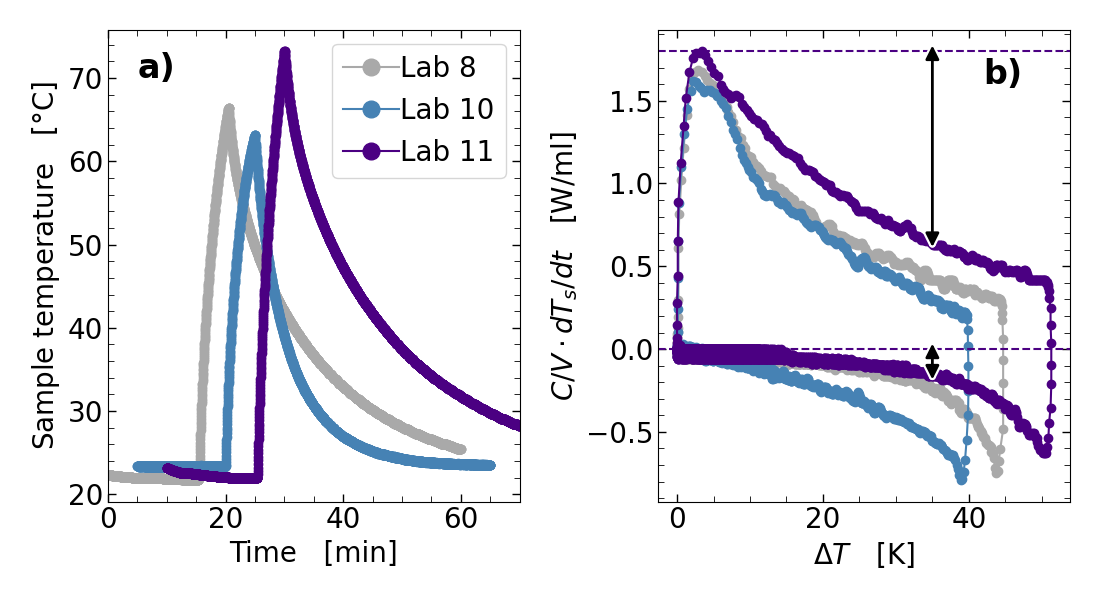}
    \caption{Data from the RADIOMAG project. All three laboratories used a field amplitude of 15 kA/m. The frequencies were 303 kHz in lab 8, 302 kHz  in lab 10, and 352 kHz in lab 11.}
    \label{fig:RADIOMAG}
\end{figure}
AC calorimetry data from 17 European state-of-the-art laboratories measuring on the same sample have been gathered by Wells {[}2021{]} as a part of the RADIOMAG EU COST action TD 1402. The data are available online {[}Wells 2020{]}. Three of the data sets are plotted in \cref{fig:RADIOMAG}a and their corresponding slope curves in \cref{fig:RADIOMAG}b. 
These are examples of 6 out of the 17 data sets from the RADIOMAG project which we have found to have a non-linear power loss with respect to $\Delta T$ (i.e. non-linear slope curves). The non-linearity can not be due to a temperature dependence of $P_\text{MNP}$ since this would require all 17 slope curves to be non-linear in the heating phase. 

It is noticeable from \cref{fig:RADIOMAG}b that the slope curves in the heating phase bend oppositely to what is expected from additional power losses (cf. \cref{fig:SlopeCurveTheory} dashed line). In fact, $P_\text{loss}$ is sub-linear with $\Delta T$. Furthermore, at any value of $\Delta T$ the power loss is higher in the heating phase than in the cooling phase (as indicated by the black arrows in \cref{fig:RADIOMAG}b). 

The remaining data sets, 
which we have not labeled \textit{"non-linear"}, are either linear (but with different $P_\text{loss}$ derived from the slope curves in the heating and cooling phases) or indecisive. 

The applicability of the corrected slope method is questionable when the power loss does not follow the linearity of Newton's law of cooling, cf. equations \eqref{eq:L}, \eqref{eq:PlossP}, and \eqref{eq:Ploss0}.
In the following, we examine the role of the sample environment on the corrected slope method.

\section{METHOD}
We conducted a set of non-adiabatic AC calorimetric measurements with different sample environments on the same commercially available sample, FeraSpin L, from Miltenyi Biotec. The sample consists of iron oxide nanoparticles with a hydrodynamic diameter of 40-50 nm dispersed in water {[}Miltenyi Biotec 2022{]}.
It had a volume of 0.8 ml, a concentration of 5.1 mg iron/ml, and was held in a 2 ml Nalgene\textsuperscript{\tiny\textregistered} cryogenic vial. Due to the low mass concentration of iron, the heat capacity of water has been used for $C$ in \eqref{eq:LumpedHeatCapacityModel}, similarly to {[}Wildeboer 2014{]}. 

\begin{figure}[t]
    \centering
    \includegraphics[width=0.9\linewidth]{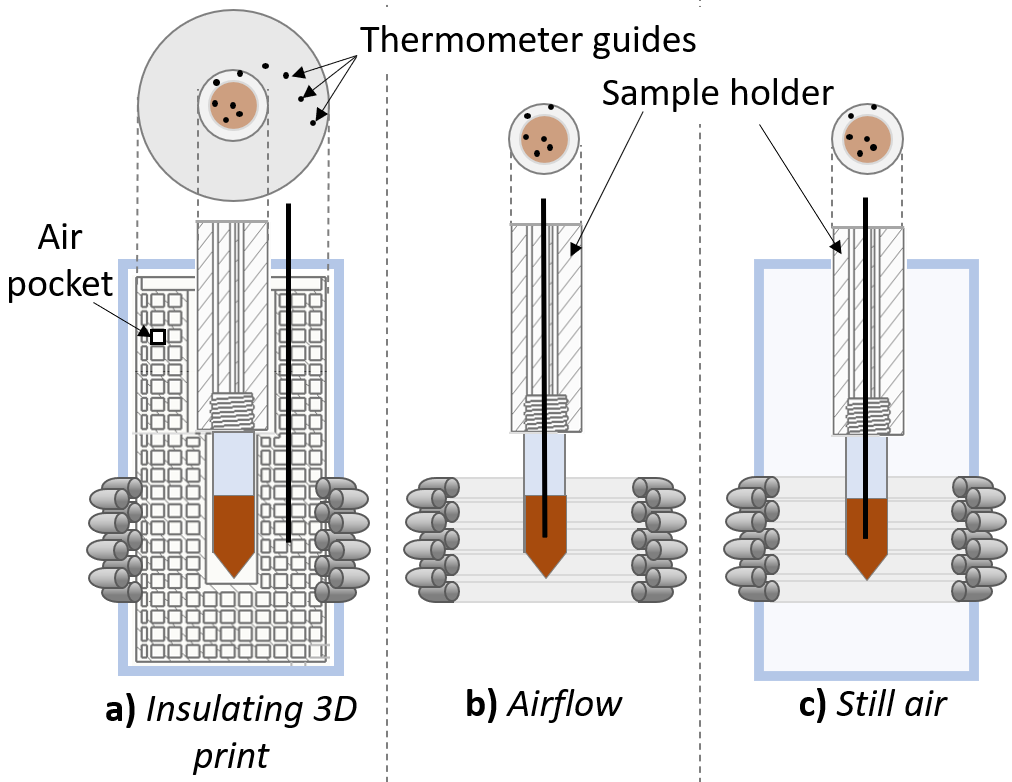}
    \caption{Illustration of the three sample environments. The black dots indicate the different radial measurement positions. The figure is not to scale.}
    \label{fig:Holder}
\end{figure}

We used a non-adiabatic AC calorimetry setup of the type MagneTherm from NanoTherics with a 9 turn water-cooled coil.
The temperature was measured by a fiber optic thermometer from OSENSA Innovations model PRB-G40. A 3D printed sample holder (\cref{fig:Holder}) was designed in-house to obtain control of the thermometer position in the horizontal and vertical directions. The sample holder has cylindrical guides for the thermometer allowing for 4 measurement positions inside the sample and 2 outside the sample. All measurements were conducted at the height of the geometric center of the sample within an uncertainty of 0.5 mm. 

Three different sample environments were used (\cref{fig:Holder}). The environment called \textit{Insulating 3D print} consists of a 3D printed insulation part. The small white squares in \cref{fig:Holder}a represent cubic air pockets with a side length of 3 mm. The \textit{insulating 3D print} has 4 additional thermometer guides outside the sample. The environment called \textit{airflow} consists of the sample mounted on the sample holder and the internal fan in the MagneTherm causes an airflow across the sample. The environment called \textit{still air} consists of the sample mounted on the sample holder and a hollow Plexiglass screens the airflow from the fan in the MagneTherm.

Before each measurement, the sample and environment were left to thermally equilibrate with the MagneTherm turned on and with the field off. The sampling period of the temperature measurements was 1 second. When calculating $dT_s/dt$ a Savitzky-Golay filter was used to smooth the data since numerical differentiation leads to a high level of noise. In this letter the Savitzky-Golay filter always has a window size of 11 seconds and a polynomial order of 2.

\section{RESULTS}
\begin{figure}[b]
    \centering
    \includegraphics[width=\linewidth]{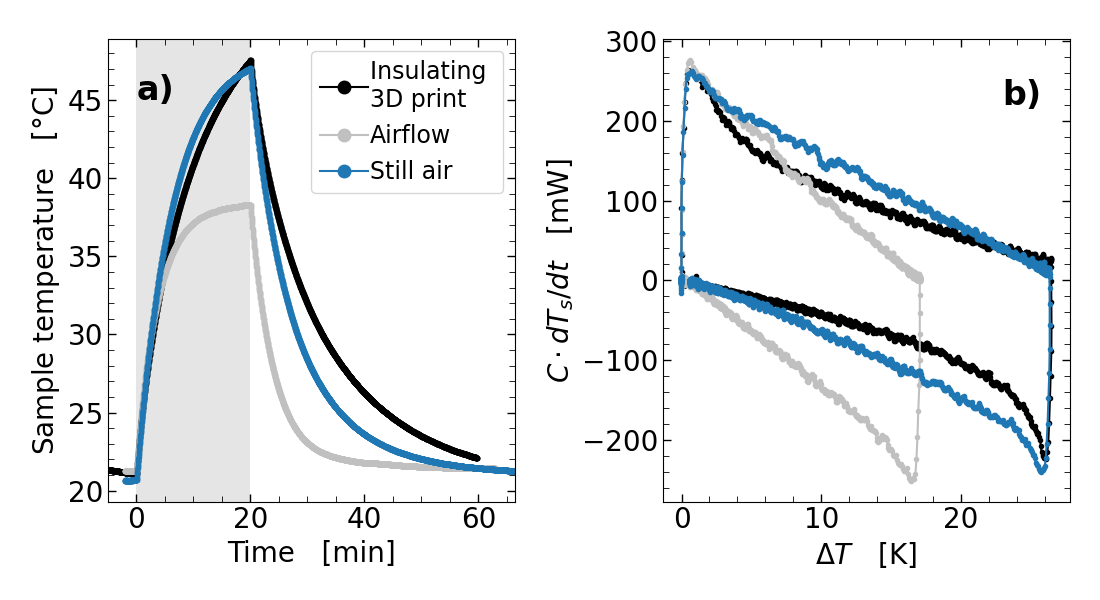}
    \caption{\textbf{a)} Temperature and \textbf{b)} slope curves where only the sample environments shown in \cref{fig:Holder} have been varied. The AC field amplitude was 4.8 kA/m and the frequency was 405 kHz.}    \label{fig:DifferentEnvironments}
\end{figure}
\cref{fig:DifferentEnvironments}a shows $T_s$ vs. $t$ measured on the same sample under the same field conditions (4.8 kA/m, 405 kHz) in the three different sample environments (\cref{fig:Holder}). The sample temperature in the \textit{insulating 3D print} and \textit{still air} environments reaches a higher value than in the \textit{airflow} due to their insulating effect. In \cref{fig:DifferentEnvironments}b the corresponding slope curves are seen. The linearity of the slope curves is found to depend on the sample environment, with the sample in \textit{insulating 3D print} being very non-linear. 
Similarly to the data shown in \cref{fig:RADIOMAG}, the slope curve for the \textit{insulating 3D print} environment is sub-linear, and the power losses extracted from the heating and cooling phases of the slope curve are different.
The non-linearity cannot be due to a temperature dependence of $P_\text{MNP}$ since this would require a non-linearity of all three slope curves. 

\begin{figure}[t]
    \centering
    \includegraphics[width=\linewidth]{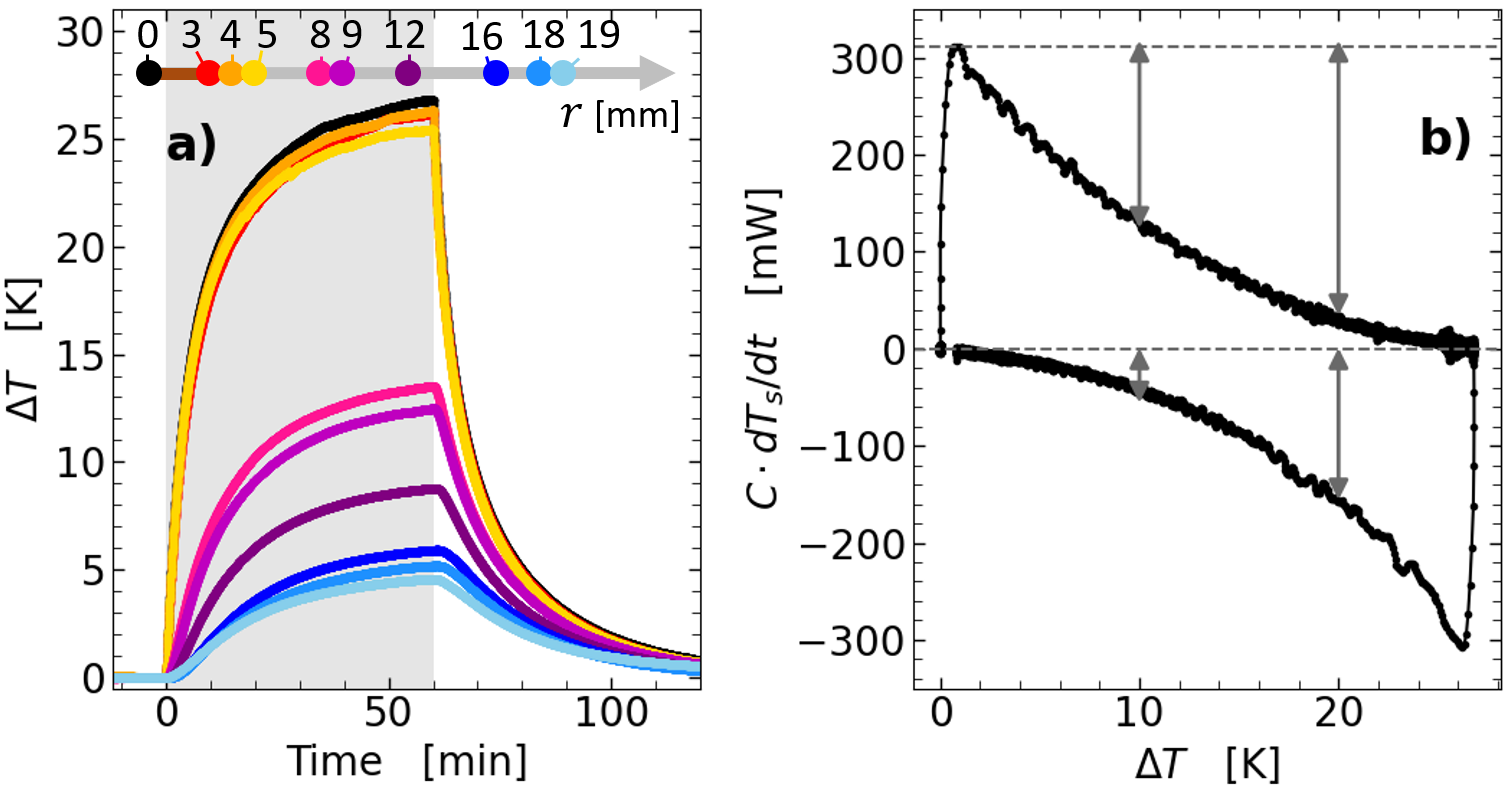}
    \caption{\textbf{a)} Temperature measurements at different radial positions, $r$. The insulation starts at $r> 5$ mm. \textbf{b)} Slope curve measured in the center of the sample. The gray arrows indicate the estimated power losses at $\Delta T$= 10 K and 20 K, assuming that $P_\text{MNP}$ is constant. The AC field amplitude was 7.2 kA/m and the frequency was 242 kHz.}
    \label{fig:TrRaw Ploss}
\end{figure}
\begin{figure}[b]
    \centering
    \includegraphics[width=\linewidth]{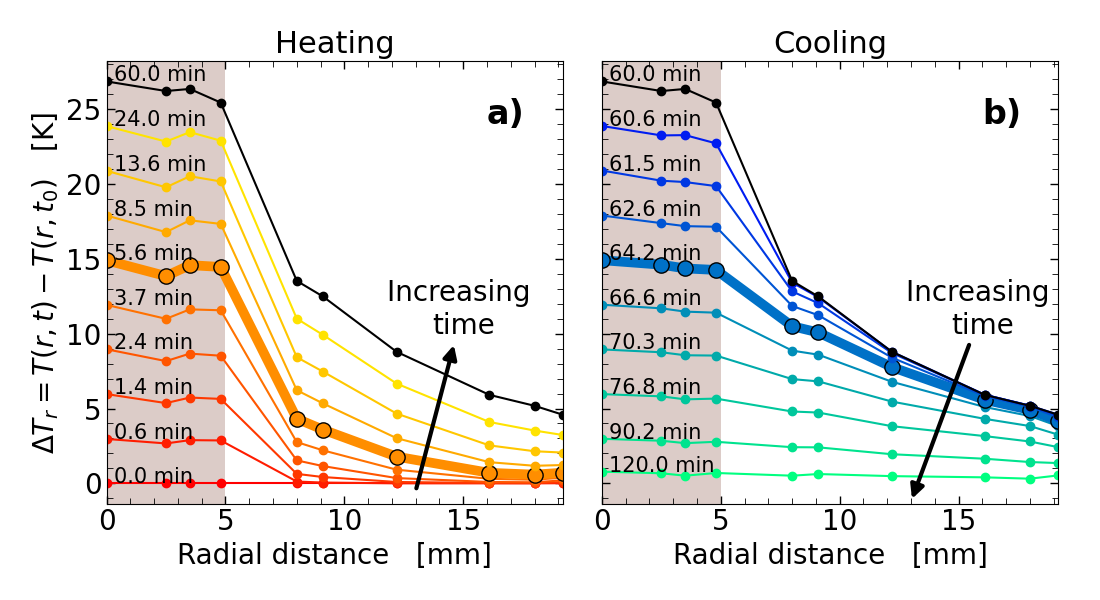}
    \caption{Temperature profiles at different times during \textbf{a)} heating and \textbf{b)} cooling phase. $t=0$ is the initiation of the field application. The brown area indicates the sample which ends at $r=5$ mm.}
    \label{fig:Tprofiles}
\end{figure}

To investigate if heating of the insulation is the cause of the non-linearity, the temperature in the sample and insulation were measured for the \textit{insulating 3D print} environment (\cref{fig:TrRaw Ploss}a). Temperature profiles were obtained from repeated heating cycles with the thermometer positioned at different radial positions and field conditions of 7.2 kA/m and 242 kHz. After each heat-up, the setup was allowed to thermally equilibrate for $>$2 hours. \cref{fig:Tprofiles} shows the temperature profiles derived from \cref{fig:TrRaw Ploss}a. 
The temperature change at each radial position $r$ is given by $\Delta T_r = T(r,t)-T(r,t_0)$. Here, $T(r,t)$ is the sample/insulation temperature,
and $t_0$ is the time when the field is switched on. 
From \cref{fig:Tprofiles} it can be seen that the temperature of the insulation 
increases significantly above the initial temperature of the system, i.e. $T(r,t>t_0)>T_0$. Consequently, the warmer insulation must support a smaller (i.e., sub-linear) power loss than that expected from the corrected slope method where $\Delta T = T_\text{s} - T_0$. The sub-linear power loss agrees with the slope curve for $T_\text{s}$ in \cref{fig:TrRaw Ploss}b. 
Moreover, it is evident from comparing the heating and the cooling profiles for the same sample temperature, 
e.g. at $\Delta T_{r=0}=\Delta T=15$ K (highlighted with fat orange and blue lines in \cref{fig:Tprofiles}), that the insulation is consistently warmer in the cooling phase than the heating phase. 
Thus, it is expected that $P_\text{loss}$ is smaller in the cooling phase since the insulation is warmer despite the 
similar value of $\Delta T$.
\cref{fig:TrRaw Ploss}b indeed shows a lower value of $P_\text{loss}$ in the cooling phase as indicated by the gray arrows.

\begin{figure}[t]
    \centering
    \includegraphics[width=0.82\linewidth]{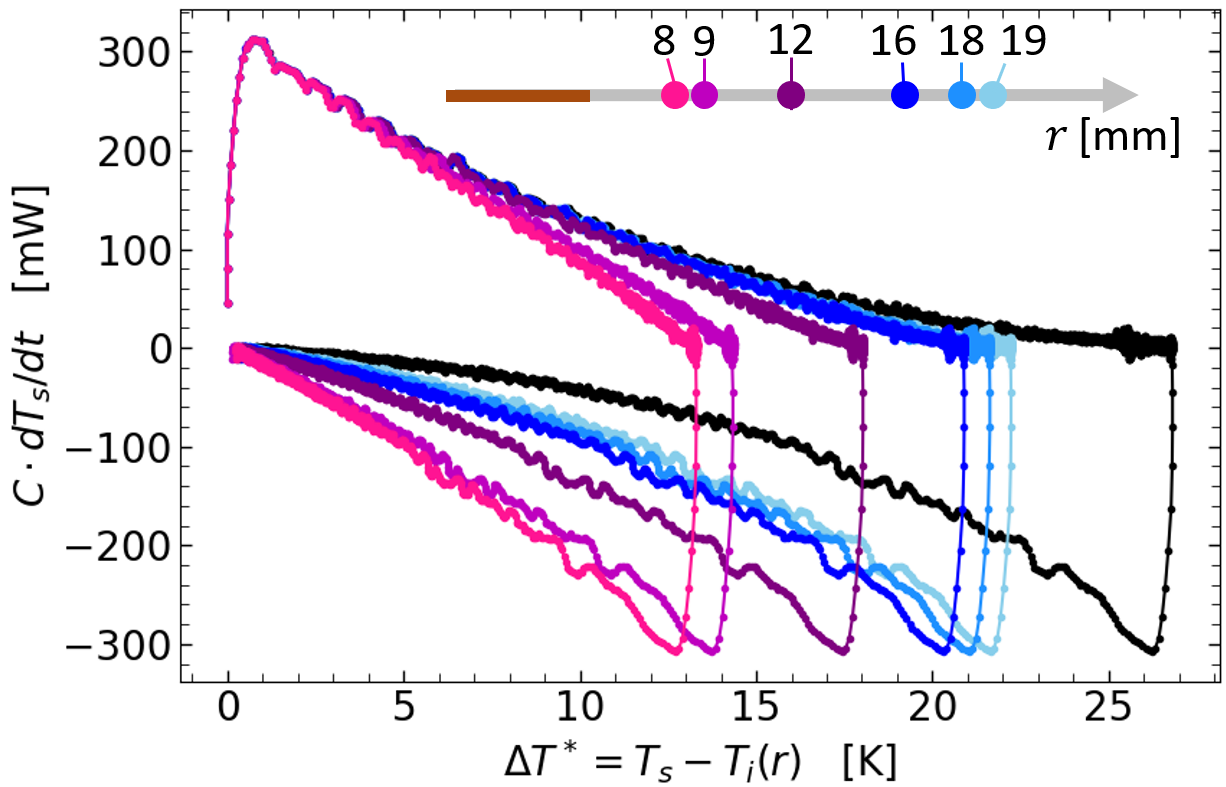}
    \caption{Slope curves with subtraction of insulation temperatures $T_i$ measured at different radial positions. The black curve is the original slope curve from \cref{fig:TrRaw Ploss}b. Each measurement of $T_i$ was done separately with the same T-probe. Small offsets, i.e. $\Delta T^*(t_0)=T_{s}(t_0) -T_{i}(t_0)\neq 0$, due to day-to-day changes in the room temperature, is subtracted. }

    \label{fig:DTstar}
\end{figure}

From \cref{fig:Tprofiles}, it follows that the initial sample temperature, $T_s(t_0)$, is not representative of the temperature of the surroundings. 
According to Newton's law, $T_0$ is the temperature of the sample surroundings and, thus, a local temperature difference seems more appropriate
\begin{align}
    \Delta T^* &= T_s-T_i.
\end{align}
Here $T_i$ is the temperature in the insulation, which depends on the radial measurement position, $r$. \cref{fig:DTstar} shows slope curves using $\Delta T^*$ for different values of $r$. The slope curves in \cref{fig:DTstar} converge towards linearity the closer $T_i$ was measured to the sample. The local temperature difference, $\Delta T^*$, captures all the causes of non-linearity. Thus, the observed non-linearity in \cref{fig:DifferentEnvironments} is ascribed purely to the temperature increase in the insulation. Based on this analysis, we propose a new method for estimating $P_\text{MNP}$ where $\Delta T$ is substituted with $\Delta T^*$ in the corrected slope method \eqref{eq:CorrectedSlope}. 
This method will require further research on the optimal positioning of the thermometer near the sample in the insulation. 

Finally, we return to \cref{fig:DifferentEnvironments}b, where the most linear slope curve was seen
for the \textit{airflow} environment. This linearity is in agreement with forced convection, where the temperature of the surroundings, $T_\text{0}$, is constantly kept equal to $T_s(t_0)$. Thus, \textit{airflow} seems suitable for permitting the original form of the corrected slope method. This might seem surprising, since insulation often is introduced to reduce $P_\text{loss}$ and improve the reliability of non-adiabatic AC calorimetry. However, when a strong cooling such as forced convection is applied, a validation of sample temperature homogeneity is needed since this is a key assumption behind \eqref{eq:LumpedHeatCapacityModel}.

In \cref{fig:DifferentEnvironments}b, it is also noticeable that there is an overlap in all three slope curves for $\Delta T\approx 0$.
We ascribe this to originate from an initially negligible temperature increase in the insulation. Despite the similarity of $dT_s/dt$ at $\Delta T \approx 0$, this does not call for the use of the initial slope method. The initial slope method is prone to underestimate $P_\text{MNP}$ since it does not extrapolate the heating power back to $\Delta T=0$ in contrast to the corrected slope method. However, a linear fit to slope curves allow to obtain $P_\text{MNP}$ from the intercept of the fit with the y-axis at $t=0$
as mentioned above. 

In conclusion, we have found that the sample environment must be considered when estimating the heating power of magnetic nanoparticles. The corrected slope method should not be rejected, but undesirable effects on temperature increments in the sample environment should be circumvented. Solutions may include substitution of $\Delta T^*$ in Newton's law of cooling, using no insulation, or using forced convection. We encourage further investigations of both the heating and the cooling phase of the slope curves as these are instrumental for understanding power losses.

\begin{ack}
The work was supported by Independent Research Fund Denmark (grant 0217-00375B), Innovation Fund Denmark (grant 5160-00004B), and the National Committee for Research Infrastructure (NUFI ESS Lighthouse programme on Quantum materials).
\end{ack}

\newpage
\end{document}